\documentclass[aps,pre,superscriptaddress,showkeys,twocolumn,longbibliography]{revtex4-2}

\usepackage{amssymb,amsmath,amsthm,mathtools,xcolor,array}
\usepackage[utf8]{inputenc}
\usepackage[T1]{fontenc}

\usepackage{siunitx}
\DeclareSIUnit\fps{fps}
\DeclareSIUnit\rpm{rpm}
\DeclareSIUnit\molar{M}
\DeclareSIUnit\OD{\text{OD$_{600}$}}
\DeclareSIUnit\cells{cells}
\DeclareSIUnit\pixels{px}
\DeclareSIUnit\rad{rad}

\definecolor{darkgreen}{rgb}{0,0.5,0.0}

\begin{document}
\title{Accumulation and depletion of \textit{E. coli} in surfaces mediated by curvature}

\author{Benjam\'{\i}n P\'erez-Estay}
\affiliation{Departamento de F\'{\i}sica, FCFM, Universidad de Chile, Santiago, Chile.}
\affiliation{Laboratoire PMMH-ESPCI Paris, PSL Research University, Sorbonne University, University Paris-Diderot, 7, Quai Saint-Bernard, Paris, France.}

\author{Mar\'{\i}a Luisa Cordero}
\affiliation{Departamento de F\'{\i}sica, FCFM, Universidad de Chile, Santiago, Chile.}

\author{N\'estor Sep\'ulveda}
\affiliation{School of Engineering and Sciences,  Universidad Adolfo Ib{\'a}{\~n}ez, Diagonal las Torres 2640, Pe\~{n}alol\'en, Santiago, Chile.}

\author{Rodrigo Soto}
\affiliation{Departamento de F\'{\i}sica, FCFM, Universidad de Chile, Santiago, Chile.}

\begin{abstract}
Can topography be used to control bacteria accumulation? We address this question in the model system of smooth-swimming and run-and-tumble \textit{Escherichia coli} swimming near a sinusoidal surface, and show that the accumulation of bacteria is determined by the characteristic curvature of the surface. For low curvatures, cells swim along the surface due to steric alignment and are ejected from the surface when they reach the peak of the sinusoid. Increasing curvature enhances this effect and reduces the density of bacteria in the curved surface. However, for curvatures larger than $\kappa^*\approx \SI{0.3}{\per \micro \meter}$, bacteria become trapped in the valleys, where they can remain for long periods of time. Minimal simulations considering only steric interactions with the surface reproduce these results and give insights into the physical mechanisms defining the critical curvature, which is found to scale with the inverse of the bacterial length. We show that for curvatures larger than $\kappa^*$, the otherwise stable alignment with the wall becomes unstable while the stable orientation is now perpendicular to the wall, thus predicting accurately the onset of trapping at the valleys.
\end{abstract}

\maketitle

\section{Introduction}

Accumulation of bacteria at surfaces is the cause of many medical and industrial concerns. Examples are infections in medical implants~\cite{Arciola2018ImplantEvasion, Costerton2005BiofilmRegulation, Veerachamy2014BacterialReview, Quirynen2002InfectiousLiterature}, contamination of medical devices such as indwelling catheters~\cite{Donlan2001BiofilmProcess}, reduction of heat, mass, and liquid transfer due to pipe biofouling~\cite{Characklis1981BioengineeringAnalysis, Mattila-Sandholm1992BiofilmReview}, and microplastic colonization in marine debris~\cite{Rummel2017ImpactsEnvironment}. Bacteria roaming their environment are attracted to surfaces due to hydrodynamic effects or chemical signals~\cite{Berke2008HydrodynamicSurfaces, Tuson2013BacteriasurfaceInteractions}, and when in contact with a surface, their adhesion is mediated by Lifshitz--van der Waals and electrostatic interactions~\cite{Jacobs2007KineticRate, Marshall1971SelectiveSeawater}. Once adhered to the surface, the irreversible process of biofilm formation can begin, in which case bacteria secrete different proteins and polysaccharides that form an extracellular matrix shared by the cells. In biofilms, bacteria are able to resist antibiotics, the host immune system, and other hostile environments, making it a serious health issue~\cite{Hall-Stoodley2004BacterialDiseases, Costerton1987BacterialDisease, Rostl2019PhightingParasites, Matz2005OffGrazing}. Thus, it is desirable to prevent bacteria adhesion while the process is still reversible.

Traditional strategies to avoid bacterial adhesion to surfaces include chemical coatings, exposition to UV light, ultrasonic vibrations, and autoclaving~\cite{Chen2013NovelInfections, Stewart2014BiophysicsInfection}. However, not all methods yield long-lasting results or can be used in any surface. Control of surface topography has risen as a practical option, as its working principle is chemical free and can be used in implants. Some topographical designs to reduce bacteria accumulation include microscopic wells of nanometric depth homogeneously distributed in space~\cite{Perera-Costa2014StudyingPatterns}, diamond-like patterns inspired in sharkskin~\cite{Reddy2011MicropatternedColi}, and hierarchically wrinkled surface topographies inspired by rose petal structures~\cite{Efimenko2009DevelopmentAntifouling, Dou2015BioinspiredAdhesion}. Typically, these patterns make the surface superhydrophobic or hydrophilic, which reduces adhesion~\cite{yuan2017surface}. Alternatively, these patterns can interrupt colony expansion by trapping the bacteria~\cite{chien2020inhibition, chung2007impact}.

Another topographical approach consists in the control of surface curvature at a larger scale of tens of micrometers. Such surfaces can be easily incorporated in medical devices exposed to blood flow or in marine vessels. The working principle lies in breaking the hydrodynamic and steric interactions that exist between swimming bacteria and solid boundaries. Steric interactions act on bacteria as they approach a wall, producing a torque that causes them to align almost parallel to the surface, with a small angle towards the wall~\cite{Bianchi2017HolographicBacteria}. Once swimming parallel to the wall, hydrodynamic interactions maintain the cells near the wall, and produce a torque that generates circular bacterial trajectories on the surface~\cite{Lauga2006SwimmingBoundaries}. This leads to the hydrodynamic trapping of bacteria on the surface, increasing their residence times close to the wall and eventually leading to adhesion~\cite{Drescher2011FluidScattering, Junot2021Run-to-tumbleBacteria}. However, highly curved convex walls can remove this trapping effect and direct bacterial motion away from the wall~\cite{Wan2008RectificationBarriers, Reichhardt2017RatchetSystems}. For instance, a critical radius was determined for cylindrical pillars, below which bacteria significantly reduce their residence time around the pillars~\cite{Sipos2015HydrodynamicWalls}. The same principle has been used to direct and concentrate swimming bacteria~\cite{Galajda2007WallOfFunnels}. 

Sinusoidal surfaces are a model geometrical pattern to study the possibilities of topography to control bacterial accumulation. In the valleys of a sinusoidally-shaped surface, the cells reorient themselves and are guided towards the peaks. There, they detach from the wall due to its curvature. It has been proven that accumulation of \textit{Escherichia coli} on this kind of surface can reduce up to \SI{50}{\percent} when compared to a flat surface thanks to this phenomenon~\cite{Mok2019GeometricAccumulation}. However, we still lack an understanding of what defines the optimal surface for the control of bacteria accumulation. In this work we characterize the dynamics of the bacteria interacting with a sinusoidal surface to determine the geometrical variables that control bacteria accumulation. Interestingly, we find three different accumulation regimes and show that the principal observables depend mainly on the characteristic curvature of the surface, showing a non-monotonic behavior. Also, in contrast to the aforementioned work, we studied both smooth swimming and tumbling bacteria, finding that tumbling affects only marginally the results. Simulations of a minimal model show that the bacterial accumulation can be quantitively and qualitatively well described considering only the steric interactions with the surface. Using these results, we present a dynamical analysis with which we accurately predict the critical curvature where accumulation starts.

\section{Experimental setup}\label{sec.setup}

\begin{figure}
    \centering
    \includegraphics[width=\linewidth]{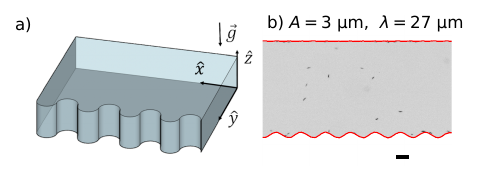}
    \caption{a) Diagram of a section of the microfluidic device with constant amplitude and wavelength. b) Snapshot obtained in a device with $A=\SI{3}{\micro\meter}$ and $\lambda=\SI{27}{\micro\meter}$. Color has been inverted, hence bacteria appear dark. The red lines represent the vertical walls of the channel and the scale bar is \SI{10}{\micro\meter}.}
    \label{channel_diagram}
\end{figure}

We use two strains of fluorescent \textit{E.\ coli} bacteria: JEK1036, which performs usual run and tumbles (``run-and-tumblers'', R\&T), and JEK1038, which is mutated to suppress tumble (``smooth swimmers'', SS). Each strain is used separately, suspended in motility buffer in dilute conditions (optical density at \SI{600}{\nano\meter} OD$_{600} = 5 \times 10^{-4}$, corresponding to, approximately, \SI{5e5}{bact/\milli\liter}). The suspensions are injected into long, \SI{100}{\micro\meter}-wide and \SI{25}{\micro\meter}-deep PDMS microfluidic channels, fabricated with standard soft lithography techniques~\cite{McDonald2002PolyDevices}. The microfluidic devices have three flat walls and a fourth undulated one in a sinusoidal form, $A \sin(2\pi x/\lambda)$. Each microchannel has sections of length \SI{1}{\milli\meter} with multiple sinusoidal periods of constant amplitude $A$ and wavelength $\lambda$, as shown in Fig.~\ref{channel_diagram}a. We work with 16 combinations of parameters $(A, \lambda)$, with nominal values $A =\SIlist[list-units=single, list-final-separator = {, }]{3;6;9;12}{\micro\meter}$ and $\lambda=\SIlist[list-units=single, list-final-separator = {, }]{21;24;27;30}{\micro\meter}$. The actual amplitude and wavelength of the sinusoidal walls can differ up to \SI{17}{\percent} from the nominal values due to imperfections of the microfabrication procedure. For each case, we measure the real amplitude and wavelength using fluorescence microscopy.

Prior to the injection, the channels are filled with $0.1\%$ BSA solution dissolved in motility buffer to prevent cell adhesion to walls. After inoculation, the inlets of the channel are sealed with glass coverslips to prevent residual flow. We image the bacteria in fluorescence using an inverted microscope (Nikon TS100F) with a 40x/0.6 NA Plan Fluor objective. Videos are recorded for two minutes at \SI{10}{\fps} with a digital camera (Andor Zyla 4.2). The focal plane is located at the bottom of the channel ($z=0$). Figure \ref{channel_diagram}b shows a typical frame obtained in the experiments. Bacteria are tracked in two dimensions using the TrackMate plugin of Fiji \cite{Tinevez2017TrackMate:Tracking}. 

\section{Experimental results} \label{sec.results}

\subsection{Global and local accumulation of bacteria at the curved walls}

\begin{figure}
    \centering
    \includegraphics[width=\linewidth]{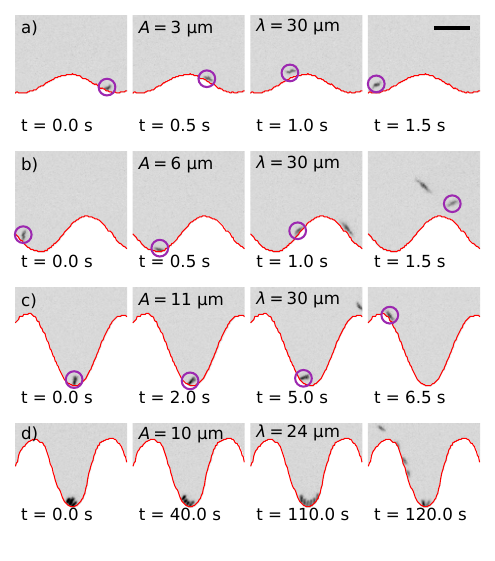}
    \caption{Color-inverted snapshot sequences obtained with fluorescence microscopy, showing smooth swimmers as they move in contact with curved walls of different amplitude $A$ and wavelength $\lambda$, with increasing curvature from top to bottom. Times indicated at the bottom of each frame are with respect to the first contact of the bacterium with the wall, except for (d), in which the cluster existed already at the beginning of the acquisition. The red lines mark the measured position of the walls and purple circles are used to highlight the bacterium that is being observed. The scale bar in the top right represents \SI{10}{\micro\meter}. Corresponding movies S1 to S4 are available in the supplementary material.}
    \label{phenomenology}
\end{figure}

To study the interaction and accumulation of bacteria at curved surfaces, we proceed to perform experiments at low concentration, allowing us to follow the motion of individual swimmers. When swimming near a wall with low curvature, smooth swimmers can follow the sinusoidal surface due to hydrodynamic attraction~\cite{Sipos2015HydrodynamicWalls, Berke2008HydrodynamicSurfaces} and steric alignment, much like near a flat wall. An example is shown in Fig.~\ref{phenomenology}a and supplementary video S1. As curvature increases, this hydrodynamic effect cannot compensate the tendency of the swimmers to continue in a straight line, making it more likely that bacteria leave the surface once they reach a peak of the sinusoidal wall. This kind of trajectories can be observed in Fig.~\ref{phenomenology}b and supplementary video S2.

For walls with even higher curvature, alignment is insufficient to rotate the swimmers when they reach a valley, resulting in bacteria swimming nearly perpendicular to the surface. This effect causes bacteria to stay in the valleys for several seconds before, eventually, escaping thanks to fluctuations, as can be seen in Fig.~\ref{phenomenology}c and supplementary video S3. 

The time that bacteria remain trapped in the valleys increases dramatically as curvature increases further, causing the arrival of multiple bacteria at the same valley. Bacteria-bacteria interactions suppress the alignment with the surface, and clusters of bacteria can remain for several minutes. Clusters formed in this manner exhibit a stable aligned state. An example is shown in Fig.~\ref{phenomenology}d and supplementary video S4. The dispersion of the cluster normally follows the destabilization caused by swimming noise or the arrival of new bacteria.

Based on these observations, we measure bacterial accumulation at the curved walls. For this, we select the bacteria that are in contact with either the curved or flat surface, where the contact bands are defined as a region \SI{4}{\micro\meter} from each surface in the $\pm y$ direction for the flat and curved wall, respectively (see Fig.~\ref{channel_diagram}a for the definition of the coordinate system). From the tracking of this subset, the bacterial density along the curved and flat surfaces, respectively $\mu_\text{curved}^\prime(x)$ and $\mu_\text{flat}^\prime(x)$ are obtained. To reduce the statistical error, we average $\mu_\text{curved}^\prime$ over the different oscillation periods and collapse it on the interval $-\lambda/2$ to $\lambda/2$, centered around the valley. Also, as the global bacterial concentration varies between experimental realizations, for each experiment we normalize the mean bacterial density on the curved surface to the spatial average of the flat one: $\mu(x) = \mu_\text{curved}^\prime(x)/\langle \mu_\text{flat}^\prime(x) \rangle_x$. Therefore, values of $\mu(x)$ larger than unity indicate an accumulation of bacteria at the curved wall in comparison with the flat one.

\begin{figure}
    \centering
    \includegraphics[width=\linewidth]{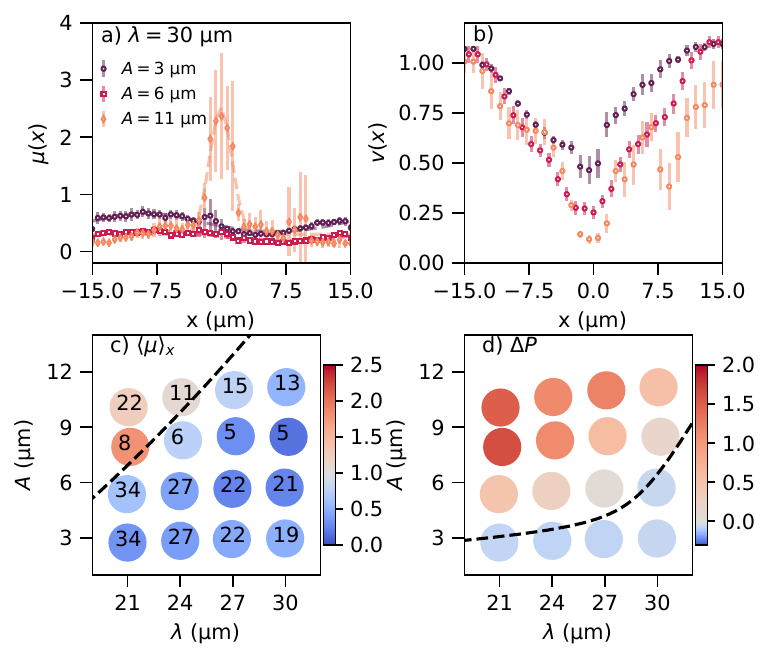}
    \caption{Normalized linear density (a) and normalized average speed of swimmers in contact with the curved walls (b) as a function of $x$ for three cases with constant wavelength $\lambda = \SI{30}{\micro\meter}$ and varying amplitude $A$. Error bars represent the \SI{95}{\percent} confidence interval for the mean. c) Spatial average of the normalized concentration profile as a function of the measured amplitudes and wavelengths, $A$ and $\lambda$. The number of surface periods considered in the averages is indicated above each data point. d) Excess accumulation at the valleys, $\Delta P$, as a function of the measured amplitudes and wavelengths, $A$ and $\lambda$. The dashed black curves show $\langle \mu \rangle_x=1$ and $\Delta P=0$, obtained from fitting the measurements to a second order polynomial in $A$ and $\lambda$. For $\Delta P$ the fit is performed only with data points near the region of interest.}
    \label{mu_v_C1}
\end{figure}

\begin{figure}
    \centering
    \includegraphics[width=\linewidth]{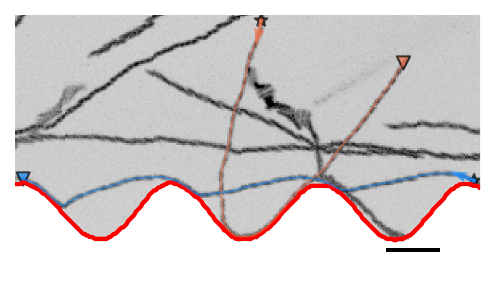}
    \caption{Examples of bacteria trajectories. The initial and final positions are represented by a star and a triangle, respectively. Due to the counterclockwise rotation at the surface, the orange trajectory leaves the curved surface at the right of the valley, so it curves out, while the opposite happens several times to the blue trajectory. The red lines represent the measured walls of the channel, and the scale bar is \SI{10}{\micro\meter}. The background is a superposition of color-inverted frames in a time interval of \SI{5}{\second}. The corresponding movie S5 is available in the supplementary material.}
    \label{trajectory_asymmetries}
\end{figure}

Figure~\ref{mu_v_C1}a presents the normalized density profiles $\mu(x)$ for the cases shown in Figs.~\ref{phenomenology}a, b, and c, which have equal wavelength but increasing amplitudes. Both for $A=\SI{3}{\micro\meter}$ and $A=\SI{6}{\micro\meter}$, the bacterial density is roughly flat, with an average value smaller than unity that decreases with the amplitude, meaning that bacteria accumulate less in the curved wall with $A = \SI{6}{\micro\meter}$ than in the one with $A = \SI{3}{\micro\meter}$. This can be understood by the ejection of bacteria at the peaks from the most curved wall, as shown in Fig.~\ref{phenomenology}b. Conversely, for $A=\SI{11}{\micro\meter}$, a maximum at $x=0$ appears, consistent with the bacteria accumulation at the valley observed in this case (Fig.~\ref{phenomenology}c). The small asymmetry of the profiles with respect to the center of the valleys at $x=0$, for $A=\SI{3}{\micro\meter}$ and \SI{6}{\micro\meter}, with more bacteria on the left half of the period, is due to the counterclockwise circular motion of bacteria when swimming near the bottom surface~\cite{Lauga2006SwimmingBoundaries}. Indeed, bacteria swimming from right to left leave the surface on the peak at the left of the valley and return more easily to the surface thanks to the counterclockwise circular trajectory, naturally increasing the density on the left side of the profile. On the contrary, bacteria leaving on the peak at the right of the valley will be directed away from the surface (see Fig.~\ref{trajectory_asymmetries}).

The average velocity on the curved side as a function of the position on the period was also measured from the tracking of bacteria in the contact bands. To reduce the deviations produced by the variability in bacterial activity from experiment to experiment, the average velocities were normalized by the average speed of bacteria in the bulk for each experiment (typically about \SI{20}{\micro\meter\per\second}). Figure~\ref{mu_v_C1}b shows the normalized velocity profiles, $v(x)$, for the same cases of Fig.~\ref{mu_v_C1}a and Figs.~\ref{phenomenology}a--c. Consistently with the observed behavior, the velocity is smaller near the valley, with decreasing values as the amplitude increases.

The global relative accumulation of bacteria at the curved wall was defined as the spatial average of the normalized density profiles, $\langle \mu \rangle_x$. Figure~\ref{mu_v_C1}c shows $\langle \mu \rangle_x$ as a function of $A$ and $\lambda$. Values smaller than one, which happen for most of the cases, mean that less global accumulation on the curved surface occurs compared to the flat one. The transition curve $\langle \mu \rangle_x = 1$ was determined from a quadratic fit to the experimental data, and is shown as the dashed line in Fig.~\ref{mu_v_C1}c. The transition occurs for amplitudes $A > \SI{6}{\micro\meter}$ in the range of wavelengths considered, and it is outside of the experimental set for $\lambda > \SI{27}{\micro\meter}$. Indeed, in most of the studied parameter space, $\langle \mu \rangle_x$ is smaller than one. There is a minimum at $A=\SI{9}{\micro\meter}$ and $\lambda=\SI{30}{\micro\meter}$, with $\langle \mu \rangle_x=0.2$, considerably smaller than one. Cases with evident accumulation at the valleys, such as the example of Fig.~\ref{phenomenology}c, still yield an average global bacterial density $\langle \mu \rangle_x < 1$, which shows that this metric does not reflect completely the onset of accumulation in the curved walls.

The preferential bacterial accumulation at the valleys is characterized by the probability of finding bacteria there. For that, from $\mu(x)$ we first obtain the probability distribution function $\hat\mu(\phi)$ of the phase $\phi=2\pi x/\lambda$, normalized such that if the distribution of swimmers were uniform, $\hat\mu(\phi)=1/(2\pi)$. Then, the pdf is fitted to a  von Mises-like distribution $A+Be^{\chi\cos(\phi-\phi_0)}$, where the degeneracy for small $\chi$ is eliminated by imposing that $\phi_0$ is in the interval $[-\pi/2,\pi/2]$ (near the valley). With the fit, the excess accumulation at the valleys is computed as the pdf at $\phi_0$ compared to the uniform case, that is,
\begin{equation}
    \Delta P \equiv  A+B e^{\chi} - \frac{1}{2\pi}.
\end{equation}
Figure~\ref{mu_v_C1}d presents $\Delta P$ as a function of $A$ and $\lambda$. The dashed curve represents $\Delta P=0$, obtained by a quadratic fit of the data, separating the regions where there is an excess of bacteria in the valleys ($\Delta P>0$) to those where is a deficit ($\Delta P<0$). 

Comparing Figs.~\ref{mu_v_C1}c and d, three regions are clearly identified. For large curvatures (small wavelength and large amplitudes), $\langle \mu \rangle_x > 1$ and $\Delta P>0$, meaning that on average more bacteria accumulate on the undulated wall, with preference for the valleys. In the opposite limit of small curvatures (large wavelengths and small amplitudes), $\langle \mu \rangle_x < 1$ and $\Delta P<0$, implying that on average the undulated wall captures less bacteria and that they preferentially accumulate at the peaks. In the intermediate region, the average accumulation on the undulated wall is smaller than on the flat one but, nevertheless, at the valleys there is an increasing concentration of bacteria. 

\subsection{Effect of curvature}

The two effects that explain the different behavior of bacteria swimming near the walls, i.e., bacteria alignment to the wall and bacteria trapping at the valleys, depend on the curvature of the wall. Starting from a flat concentration profile for a flat wall, the accumulation of bacteria at the walls decreases with increasing curvatures as bacteria leave more easily the surfaces at the peaks. For larger curvatures, bacteria get trapped in the valleys, increasing the accumulation for the higher curvatures. This non-monotonic dependence suggests that a minimum of accumulation should occur for a critical value of curvature.

\begin{figure*}
	\centering
	\includegraphics[width=\linewidth]{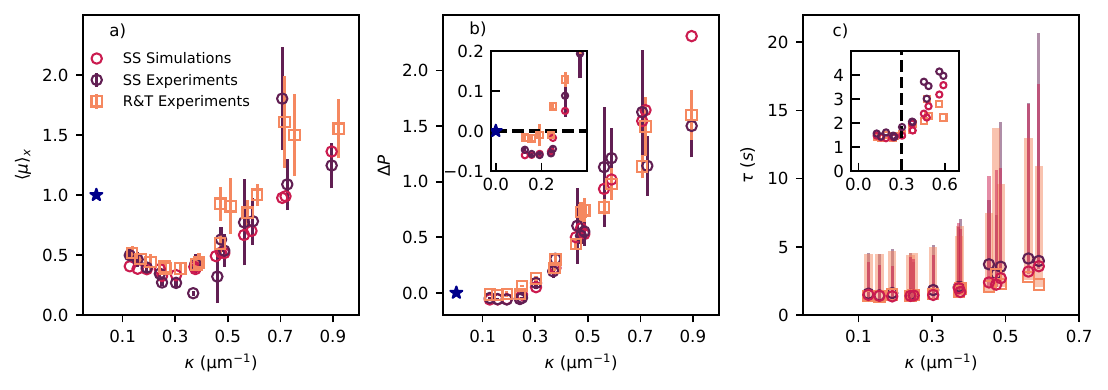}
	\caption{a)Average accumulation at the curved wall $\langle \mu \rangle_x$, b) excess accumulation at the valleys $\Delta P$, and c) residence time $\tau$ as a function of the maximum curvature of the walls, $\kappa = 4\pi^2A/\lambda^2$. The insets show a zoom of the small curvature region. As a reference, the theoretical values for vanishing curvature $\langle \mu \rangle_x = 1$, $\Delta P = 0$ are shown with a star in (a) and (b). For $\langle \mu \rangle_x$ and $\Delta P$ error bars represent the $95\%$ confidence interval for the average of the parameter. For the contact times, they represent the interval that contains $90\% $ of the measured contact times. Contact times for curvatures $\kappa > \SI{0.7}{\per\micro\meter}$ are immeasurable due to the formation of clusters.}
	\label{meanI_and_c1}
\end{figure*}
  
Since the valleys and the peaks are key to determining the dynamics of bacteria, it is natural to study the system with respect to the curvature of the wall in those positions, given by $\kappa = 4\pi^2 A / \lambda^2$. Figures~\ref{meanI_and_c1}a and b show the average normalized density $\langle \mu \rangle_x$ and the excess accumulation at the valleys $\Delta P$ as functions of $\kappa$. The collapse of data shows that bacterial accumulation on curved surfaces is primarily determined by the maximum curvature of the wall. In the limit of vanishing curvature, the flat surface is recovered, implying that there one should get $\langle \mu \rangle_x=1$ and $\Delta P=0$, which are shown with a star in the figures. The average density presents the non-monotonic behavior described before, with a minimum accumulation for $\kappa_\text{min} \approx\SI{0.4}{\per\micro\meter}$. Only for curvatures larger than \SI{0.7}{\per\micro\meter}, the average accumulation on the curved wall becomes larger than the flat one.

The excess accumulation at the valleys $\Delta P$ shows a sign change at a critical curvature. Negative values are measured for small curvatures, meaning that in those cases the maximum accumulation, although weak, occurs at the peaks. Conversely, $\Delta P$ is positive and with larger absolute values for large curvatures, corresponding to preferential accumulation at the valleys. The transition takes place for $\kappa^* \approx\SI{0.3}{\per\micro\meter}$.

The residence time of bacteria in the contact zones of the curved surfaces was also measured from bacterial tracking. Leaving contact was defined as being away from the contact zones for longer than \SI{0.5}{\second}, that is, not returning to the surface before that time. The average contact times  $\tau$ are shown in Fig.~\ref{meanI_and_c1}c. Although there is a large dispersion, the average residence times remains constant around \SI{1.5}{\second} for smaller curvatures than the critical value $\kappa^*$. For larger curvatures, the average of $\tau$ increases as bacteria start to be trapped in the valleys. Contact times for curvatures smaller than $\kappa^*$ are one order of magnitude smaller than what is observed for flat surfaces \cite{Drescher2011FluidScattering}.

\subsection{Run-and-tumble swimmers}

The behavior of R\&T swimmers is qualitatively similar to SS and quantitatively only slight changes are observed. Figure~\ref{meanI_and_c1} includes the mean normalized bacterial density $\langle \mu \rangle_x$, the excess accumulation at the valley $\Delta P$, and the contact times $\tau$ as a function of the maximum wall curvature $\kappa$ for R\&T. The average normalized density $\langle \mu \rangle_x$ for R\&T swimmers (Fig.~\ref{meanI_and_c1}a) retains its non-monotonic dependence with the curvature. In comparison with SS, $\langle \mu \rangle_x$ for R\&T swimmers is larger, because tumbles facilitate escaping from the flat wall, affecting therefore the normalization. Although this effect also helps bacteria escaping from the curved wall, ejection at the peaks is dominant in this case and tumbling does not appreciably change the bacterial density at the curved wall. The minimum accumulation takes place at a similar curvature $\kappa_\text{min} \approx \SI{0.3}{\per\micro\meter}$ than for the SS.

The value of $\Delta P$ for R\&T (Fig.~\ref{meanI_and_c1}b) is also negative for low curvatures and transitions to positive values for the same $\kappa^*$. In comparison with SS, the values of $\Delta P$ for R\&T after the transition are larger (i.e.\ less negative). This is because, while the ejection of bacteria at the peaks, together with the swimming persistence, prevented SS to reach the valleys, in this case R\&T can approach the walls in any position, thus making $\Delta P$ closer to 0.

The residence time (Fig.~\ref{meanI_and_c1}c) behaves similarly as for SS, with large dispersion but a relatively constant average value at around \SI{1.5}{\second} for low curvatures and starting to increase when bacteria begin to be trapped. Nevertheless, the average values are smaller because tumbling facilitates bacteria escaping from the walls. In any case, the effects of the valley to trap bacteria are clear when looking at $\langle\mu\rangle_x$ and $\tau$, revealing that the tumbling of bacteria does not alter either the capacity of the surface to remove bacteria at low curvature or the critical curvature for the onset of accumulation at the valleys.

\section{Simulations}\label{sec.simulations}

In order to capture the essential dynamics of the system, we simulate active Brownian particles (ABP) moving in two dimensions~\cite{PhysRevE.98.052141}, with periodic boundary conditions on the lateral directions and confined by a flat and a sinusoidal wall. By doing so, we eliminate the possibility for bacteria to approach from and escape to the third dimension. In the simulations, we neglect the circular motion of the swimmers in contact with the bottom and upper walls~\cite{Lauga2006SwimmingBoundaries}, which although gives rise to the small asymmetry on the density profiles, is subdominant for the reported phenomena, as well as translational noise, which is negligible compared to self-propulsion.

Each swimmer is described by its position $\mathbf{r}$ and director $\hat{\mathbf p}=(\cos\theta,\sin\theta)$, where the director angle $\theta$ is subject to rotational noise of intensity $D_\text{r}$. The swimmers are modeled as disks with radius $L_\text{cm}$. We focus on the dilute regime, so particle interactions, both hydrodynamic and steric, are negligible. Swimmers do, however, interact with the walls. The total force on each swimmer, which vanishes in the low-Reynolds regime, is $\gamma u \hat{\mathbf p}+F_\text{wall}\hat{\mathbf n}_\text{wall}$, where $\gamma$ is the viscous drag coefficient, $u$ is the self-propulsion speed of the swimmers, and $\hat{\mathbf n}_\text{wall}$ the local unit vector normal to the wall, pointing into the swimming domain. The force exerted by the wall when swimmers are in contact with it is obtained by imposing $\dot{\mathbf{r}} \cdot \hat{\mathbf n}_\text{wall} = 0$, resulting in  $F_\text{wall}=- \gamma u \hat{\mathbf p} \cdot \hat{\mathbf n}_\text{wall}$. 

When bacteria are in contact with a wall, hydrodynamic and steric interactions induce an alignment of the swimmer to the surface~\cite{Sipos2015HydrodynamicWalls, Bianchi2017HolographicBacteria}. In simulations, this effect is modeled as a torque, $\tau_\text{wall}$, associated with the force from the wall, which we assumed is applied in the bacterium at a position $L_\text{cm}\hat{\mathbf{p}}$ with respect to the swimmer center of mass. This gives $\tau_\text{wall} = (L_\text{cm}\hat{\mathbf{p}} \times  F_\text{wall}\hat{\mathbf{n}}_\text{wall}) \cdot \hat{\mathbf{z}} = - u L_\text{cm} \gamma (\hat{\mathbf{p}} \cdot \hat{\mathbf{n}}_\text{wall} ) (\hat{\mathbf{p}} \cdot \hat{\mathbf{t}}_\text{wall} )$, where $\hat{\mathbf{t}}_\text{wall}=\hat{\mathbf{n}}_\text{wall}\times\hat{\mathbf{z}}$ is the tangent to the wall.

In summary, the equations of motion for each smooth swimmer are
\begin{align}
    \dot {\mathbf{r}} &= u \hat{\mathbf{p}} - u (\hat{\mathbf{p}} \cdot \hat{\mathbf{n}}_\text{wall}) \hat{\mathbf{n}}_\text{wall} \Gamma(\mathbf{r}, \hat{\mathbf{p}}), \\
    \dot\theta &= -  (\hat{\mathbf{p}} \cdot \hat{\mathbf{n}}_\text{wall} ) (\hat{\mathbf{p}} \cdot \hat{\mathbf{t}}_\text{wall} ) \Gamma(\mathbf{r}, \hat{\mathbf{p}})/T + \sqrt{2D_\text{r}} \eta(t),
    \label{dot_theta_eq}
\end{align}
where $T = (u L_\text{cm} \gamma/\gamma_\text{r})^{-1}$ is the steric reorientation time, with $\gamma_\text{r}$ the rotational drag coefficient, $\eta$ is a white noise, and $\Gamma(\mathbf{r}, \hat{\mathbf{p}})$ is a Heaviside step function, equal to one if the particle is in contact with the surface and swimming towards it. The equations are integrated with the forward Euler scheme.

Based on the experimental results we use $u = \SI{20}{\micro\meter\per\second}$ and $L_\text{cm} = \SI{0.5}{\micro\meter}$. Given that the bacteria of the experiment resemble elongated spherocylinders instead of perfect spheres, the two parameters $D_\text{r}$ and $T$ were fitted such as to better reproduce in the simulations the experimental observable  $\Delta P$ for SS (Fig.~\ref{mu_v_C1}d). The optimal values thus found are $T = \SI{0.25}{\second}$ and $D_\text{r} = \SI{0.04}{\square\rad \per \second}$. With these values, the model adequately reproduces the average density $\langle \mu \rangle_x$, the excess valley accumulation $\Delta P$, and the average contact times $\tau$, as shown in Fig.~\ref{meanI_and_c1}.

\begin{figure}
    \centering
    \includegraphics[width=\linewidth]{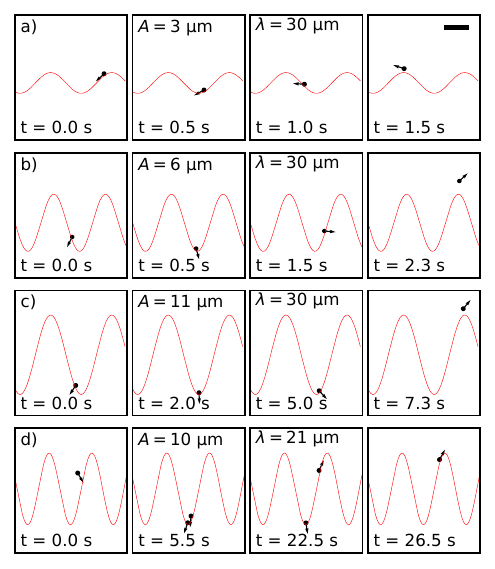}
    \caption{Simulation snapshots, analog to Fig.~\ref{phenomenology}, showing smooth swimmers moving in contact with the surface. Times indicated at the bottom of each frame are with respect to the first contact of the particle with the wall. The red lines mark the position of the walls and the scale bar is equivalent to \SI{10}{\micro\meter}. Corresponding movies S6 to S9 are available in the supplementary material.}
    \label{simulations}
\end{figure}

The model does not only reproduce the macroscopic properties, but also the dynamics of individual swimmers. Figure~\ref{simulations} shows simulated swimmer trajectories for a wall with curvature $\kappa < \kappa_\text{min}$ (Fig.~\ref{simulations}a and supplementary video S6), $\kappa_\textrm{min} < \kappa < \kappa^*$ (Fig.~\ref{simulations}b and supplementary video S7), and $\kappa > \kappa^*$ (Figs.~\ref{simulations}c and d and supplementary videos S8 and S9). In the latter, the absence of swimmer-swimmer interactions prevents the formation of clusters of trapped swimmers.

In the case of including tumbling, a new director angle is chosen at random with a rate $\nu_\text{tumble}$. Here, a third parameter is added to the model, which is fitted as for SS to best reproduce the experimental measurements of $\Delta P$, obtaining $\nu_\text{tumble}=1/\SI{7.0}{\second}$. The results  of the simulation for the average accumulation, the excess accumulation at the valleys, and the residence times as a function of the curvature (not shown to avoid overcrowding the plots) are in very good agreement with the experiments and, as for the experiments, do no differ appreciably to the results for SS. 

The successful comparison with the experiments shows that the observed phenomena we report are mainly due to the steric interactions of swimmers with the walls. That is, we were able to reproduce qualitatively and quantitatively the observed phenomena without appealing to hydrodynamic attraction to the walls or to the velocity reduction due to lubrication effects.

\section{Critical curvature}\label{sec.curvature}

The simulations show that the accumulation of bacteria at curved surfaces can be well described by a simple model where the swimmers self-propel and interact sterically with the surface. This interaction is the responsible for aligning bacteria to the surface. With this model in mind, we can compute the critical curvature as follows. Consider for simplicity that the curvature $\kappa$ of the wall is fixed and that the swimmer with director $\hat{\mathbf{p}}$ is already interacting with the surface, with $\hat{\mathbf{t}}_\text{wall}$ and $\hat{\mathbf{n}}_\text{wall}$ the local tangential and normal unit vectors to the surface. To characterize the orientation of the swimmer with respect to the surface, we define $\xi=\hat{\mathbf{p}}\cdot\hat{\mathbf{t}}_\text{wall}$, such that $\xi=\pm1$ means complete alignment with the surface and $\xi=0$ is when it is perpendicular to the surface. For its time derivative, $\dot\xi = \dot{\hat{\mathbf{p}}}\cdot\hat{\mathbf{t}}_\text{wall} + \hat{\mathbf{p}}\cdot\dot{\hat{\mathbf{t}}}_\text{wall}$, the noiseless version of Eq.~\eqref{dot_theta_eq} for a swimmer in contact with the surface dictates that $\dot{\hat{\mathbf{p}}}=\dot\theta\hat{\mathbf{p}}\times\hat{\mathbf{z}}$, with $\dot\theta=-(\hat{\mathbf{p}}\cdot\hat{\mathbf{t}}_\text{wall})(\hat{\mathbf{p}}\cdot\hat{\mathbf{n}}_\text{wall})/T$. The temporal derivative of the tangential vector is obtained from the kinematics on a curved trajectory, 
\begin{equation}
\dot{\hat{\mathbf{t}}}_\text{wall} = \frac{d\hat{\mathbf{t}}_\text{wall}}{ds}\frac{ds}{dt}= \hat{\mathbf{n}}_\text{wall}\left(u\kappa\hat{\mathbf{p}}\cdot\hat{\mathbf{t}}_\text{wall}\right),
\end{equation}
where $s$ is the arc length. Using that $\hat{\mathbf{p}}\cdot\hat{\mathbf{n}}_\text{wall}=\sqrt{1-\xi^2}$, it results
\begin{equation}
    \dot{\xi} = \xi \sqrt{1-\xi^2}\left( \sqrt{1-\xi^2} - u \kappa T\right)/T .
    \label{psi_dot}
\end{equation}
This equation shows that $\xi$ presents five fixed points. Of these, $\xi=\pm1$ are unstable for any positive curvature and can be neglected for the analysis. For $u\kappa T<1$, $\xi=0$ is unstable and $\xi=\pm\sqrt{1-(u\kappa T)^2}$ are stable fixed points, corresponding to configurations where the swimmers move along the surfaces forming an angle with respect to the surface $\theta=\sin^{-1}(u\kappa T)$, either to the right or to the left. The swimming angle increases with the curvature until $u\kappa T=1$, where it becomes perpendicular. For larger curvatures, now the perpendicular configuration (fixed point $\xi=0$) becomes stable and the other fixed points disappear. The transition takes place when the aligning torque, represented by the first term in Eq.~\eqref{psi_dot}, is not able to compensate the continuous change in the direction of the tangent vector, which is accounted for by the second term in the equation.

Hence, we have identified a critical curvature of $\kappa_\text{theo}^*=1/(uT)=L_\text{cm}\gamma/\gamma_\text{r}$, such that surfaces with higher curvatures will trap the bacteria making them swim perpendicular to the surface. The drag coefficients scale as $\gamma\sim L_\text{cm}$ and $\gamma_\text{r}\sim L_\text{cm}^3$, with  prefactors that depends on the geometry of the swimmer, implying that $\kappa_\text{theo}^*\sim L_\text{cm}^{-1}$. The critical curvature is  hence controlled by the bacterial length and its geometry. The parameters used in the simulation give $\kappa_\text{theo}^*=\SI{0.2}{\per\micro\meter}$, which is comparable to the value of $\kappa^*=\SI{0.3}{\per\micro\meter}$ obtained from the experiments.

One key aspect of the derivation is that we assumed fixed positive curvature. Therefore, swimmers moving inside circular surfaces would remain trapped if the curvature is larger than the critical. Nevertheless, in simulations and experiments with a sinusoidal wall, the curvature decreases when bacteria swim away from the valley. This, coupled with the presence of rotational diffusion leads to the possibility of exiting the trap even for curvatures larger than the critical. The balance between curvature and rotational diffusion defines the time bacteria are trapped in the valley of sinusoidal surfaces.

\section{Discussion and conclusions}\label{sec.conclusions}

Using experiments of \textit{E.\ coli} swimming in microfluidic devices with fabricated sinusoidal walls of controlled amplitudes and wavelengths, we have studied the interaction of these bacteria with curved walls and the accumulation that results. The global accumulation on the curved wall compared to the reference flat one on the opposite side of the channel, as well as the spatial distribution of swimmers on the the curved wall are controlled mainly by the maximum curvature $\kappa$ of the curved wall and not separately by the amplitude or wavelength. The dependence on the curvature is not monotonic and three regimes are identified. For low curvatures, $\kappa<\SI{0.3}{\micro\meter^{-1}}$, the curved wall captures less bacteria than the flat one and the bacteria that are close to the curved wall show a weak preference to be located close the the sinusoidal peaks. For curvatures in the range $\SI{0.3}{\micro\meter^{-1}}<\kappa<\SI{0.7}{\micro\meter^{-1}}$, the average accumulation on the curved wall is still smaller than on the flat wall, but now bacteria are preferentially located at the valleys. Finally, for curvatures larger than $\SI{0.7}{\micro\meter^{-1}}$, the average accumulation on the curved wall is larger than on the flat one and the number of bacteria in the valleys start to grow rapidly with curvature, forming long-lived clusters. These results, obtained for smooth, non-tumbling swimmers, are preserved also when studying run-and-tumble bacteria. 

The three regimes can be qualitatively well understood by following the kinematics of bacteria swimming near the walls, where we were able to identify that for low curvatures, bacteria can persistently swim parallel to the walls. When increasing the curvature, bacteria align with the surface and are ejected near the peaks. For curvatures below $\SI{0.4}{\micro\meter^{-1}}$, the contact times with the surface are smaller than \SI{2}{\second} on average, and they start to grow considerably for larger curvatures. Finally, when the curvature is too large, alignment is not possible and bacteria are trapped on the valleys which can even lead to the formation of clusters. When clusters are formed, tracking becomes difficult and it is not possible to obtain the contact times accurately, but values larger than \SI{200}{\second} were observed. These increasing contact times make it more likely for bacteria to adhere to the surface.

Although hydrodynamic interactions between the walls and the bacteria are undoubtedly present, as evidenced by the decrease of bacteria speed near the walls, we demonstrated numerically that they are not relevant to describe the described behavior. A minimal computational model was simulated, where we consider active Brownian particles moving on two dimensions, interacting only sterically with the walls, which reorient the swimmers as a result of  being impenetrable. Fitting the rotational Brownian diffusivity and the reorientation time, the model accurately captures all the observed phenomena. The success of the simulations indicates that the reorientation and escape dynamics are controlled mainly by the steric interactions. This result allowed us to build a dynamical model with which the critical curvature that controls the accumulation on the valleys is accurately predicted to be inversely proportional to the bacterial body length.

In our experiments we do not allow for bacteria to divide and, also, the treatment of the surfaces impedes bacteria to adhere. Consequently, no biofilm is form, even in cases where crowding was observed in the valleys. Nevertheless our results give important information on the early stages of bacterial accumulation: the recognition of three different regimes and the nonmonotonic dependence with curvature should be considered in designing devices aiming to control bacterial accumulation. For example, the sole quantification by the average accumulation can produce misleading results as it is possible to have less bacteria on average on the curved wall, but the formation of clusters in the valleys. Since clustering of bacteria can potentially lead to infection foci, this is key to design curved surfaces that avoid bacterial accumulation.

\section*{Acknowledgments}
This research was supported by the Millennium Science Initiative Program NCN19\_170D and Fondecyt Grants No. 1220536 (R.S. and N.S.) and 1210634 (M.L.C.), all from ANID, Chile, and the Franco--Chilean Ecos--Sud Collaborative Program 210012. Fabrication of microfluidic devices was possible thanks to ANID Fondequip grants EQM140055 and EQM180009. This project has received funding from the European Union’s Horizon 2020 research and innovation program under the Marie Skłodowska-Curie grant agreement No 955910. We also thank Eric Clement and Anke Lindner for fruitful discussions.


%

\end{document}